\documentclass[10pt,twocolumn,letterpaper]{article}
\pdfoutput=1
\usepackage[]{iccv}      % To produce the REVIEW version
\usepackage{multirow}
\usepackage{makecell}
\usepackage{array}
\usepackage{graphicx}
\newcommand{\mystrut}{\rule{0pt}{1.2em}}
\newcolumntype{P}[1]{>{\centering\arraybackslash}p{#1}}
\newcolumntype{L}[1]{>{\raggedright\arraybackslash}m{#1}}

\newtheorem{problem}{Problem}

\newcommand{\method}{{SecDOOD}\xspace}

\definecolor{iccvblue}{rgb}{0.21,0.49,0.74}
\usepackage[pagebackref,breaklinks,colorlinks,allcolors=iccvblue]{hyperref}

\def\paperID{30} % *** Enter the Paper ID here
\def\confName{ICCV}
\def\confYear{2025}

\title{Secure On-Device Video OOD Detection Without Backpropagation}

\author{
Shawn Li$^1$, Peilin Cai$^1$, Yuxiao Zhou$^2$, Zhiyu Ni$^3$, Renjie Liang$^4$, \\ 
You Qin$^2$, Nian Yi$^1$, Zhengzhong Tu$^5$, Xiyang Hu$^6$, Yue Zhao$^1$\\
$^1$University of Southern California  
\quad $^2$National University of Singapore \\  
$^3$University of California, Berkeley  
\quad$^4$University of Florida \\  
$^5$Texas A\&M University   
\quad $^6$Arizona State University \\
{\tt\small \{li.li02, peilinca, yinian, yzhao010\}@usc.edu,} \\
{\tt\small \{e1011019, e0962995\}@nus.edu.sg,} \\
{\tt\small zhiyuni@berkeley.edu, liang.renjie@ufl.edu,} \\
{\tt\small tzz@tamu.edu, xiyang.hu@asu.edu}
}

\begin{document}
\maketitle

\begin{abstract}
Out-of-Distribution (OOD) detection is critical for ensuring the reliability of machine learning models in safety-critical applications such as autonomous driving and medical diagnosis. 
While deploying personalized OOD detection directly on edge devices is desirable, it remains challenging due to large model sizes and the computational infeasibility of on-device training. Federated learning partially addresses this but still requires gradient computation and backpropagation, exceeding the capabilities of many edge devices.
To overcome these challenges, we propose \textbf{SecDOOD}, a secure cloud-device collaboration framework for efficient on-device OOD detection \textit{without} requiring device-side backpropagation.
SecDOOD utilizes cloud resources for model training while ensuring user data privacy by retaining sensitive information on-device. 
Central to SecDOOD is a HyperNetwork-based personalized parameter generation module, which adapts cloud-trained models to device-specific distributions by dynamically generating local weight adjustments, effectively combining central and local information without local fine-tuning. Additionally, our dynamic feature sampling and encryption strategy selectively encrypts only the most informative feature channels, largely reducing encryption overhead without compromising detection performance.
Extensive experiments across multiple datasets and OOD scenarios demonstrate that SecDOOD achieves performance comparable to fully fine-tuned models, enabling secure, efficient, and personalized OOD detection on resource-limited edge devices. To enhance accessibility and reproducibility, our code is publicly available at %\url{https://anonymous.4open.science/r/SecDOOD/}.
\url{https://github.com/Dystopians/SecDOOD}.
\end{abstract}    
\section{Introduction}

% Out-of-Distribution (OOD) detection \cite{oodbaseline17iclr,qin2024metaood, grahamdenoising, Li_2023_CVPR,Bai_2024_CVPR,li2024dpudynamicprototypeupdating} is a fundamental problem in machine learning, crucial for ensuring the reliability of models deployed in real-world applications \cite{cho2023training,yi2023uncertainty,xu2024lego, li2024panoptic,hao2024artificial,Du_2022_CVPR, sun2022icse,li2023biased}. OOD detection aims to distinguish in-distribution (ID) data from unknown OOD samples that exhibit semantic shifts and do not belong to predefined categories. This capability is essential for various safety-critical applications, such as autonomous driving \cite{SuperFusion,li2024light}, and medical imaging \cite{karimi2022improving}, where the failure to detect OOD samples can lead to severe consequences.

Out-of-Distribution (OOD) detection \cite{oodbaseline17iclr,qin2024metaood,grahamdenoising,Li_2023_CVPR,Bai_2024_CVPR,li2024dpudynamicprototypeupdating} is critical for ensuring the reliability of machine learning systems deployed in real-world scenarios \cite{cho2023training,yi2023uncertainty,xu2024lego, li2024panoptic,hao2024artificial,Du_2022_CVPR, sun2022icse,li2023biased}. It involves distinguishing in-distribution (ID) samples from unknown OOD samples, which exhibit semantic shifts and unexpected characteristics. 
Effective OOD detection is critical in safety-sensitive domains, e.g., autonomous driving \cite{SuperFusion,li2024light} and medical imaging \cite{karimi2022improving}, where incorrect identification can result in severe outcomes.

\begin{figure}[t]
    \centering
    \includegraphics[width=0.98\linewidth]{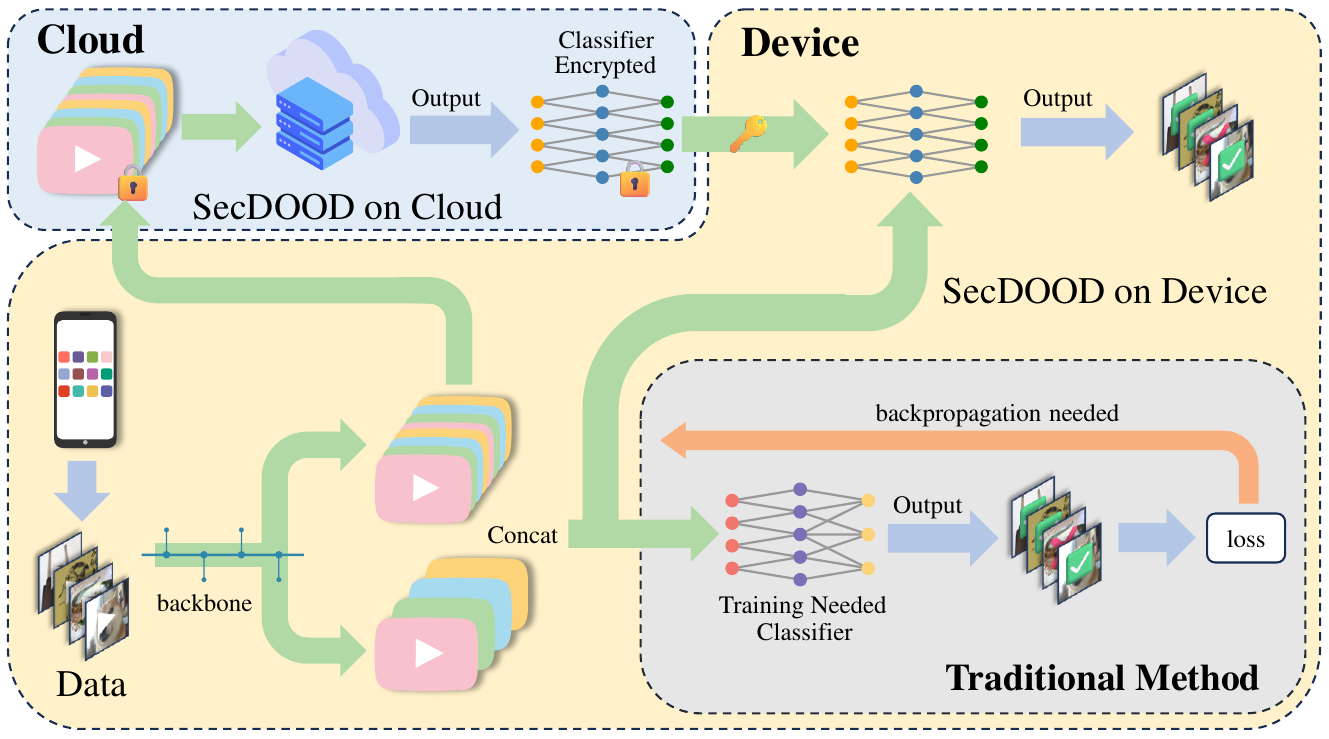} 
    \caption{Comparison of traditional OOD detection methods and our proposed \method. Conventional approaches require training or fine-tuning the classifier/backbone model for OOD detection, which is computationally demanding and challenging to execute on resource-constrained devices. In contrast, \method introduces a hypernetwork-based solution that eliminates the need for backpropagation on the device, enabling efficient and lightweight OOD detection.}
    \label{fig:compare}
    \vspace{-1em}
\end{figure}

\noindent \textbf{Need for real-time OOD Detection on Edge Devices}.
In practical scenarios, OOD detection often involves deployment directly on \textit{edge devices}, such as autonomous vehicles or medical diagnostic equipment, to accommodate user-specific data distributions and real-time detection requirements.
However, deploying OOD detection directly on edge devices faces significant practical challenges. 
\textit{First}, modern OOD detection models often have large parameter sizes \cite{oodbaseline17iclr,dong2024multiood}, making deployment infeasible on resource-limited edge hardware. 
\textit{Second}, conventional on-device training or fine-tuning is computationally prohibitive, given limited hardware capabilities. 
Federated learning \cite{collins2022fedavg,sabah2024model} addresses data privacy concerns but still requires gradient computation and backpropagation, which exceed the capacity of many edge devices.

% To address these challenges, we propose \textbf{Sec}ure On-\textbf{D}evice \textbf{OOD} Detection Framework named \textbf{SecDOOD}, a cloud-device collaboration framework for secure and efficient OOD detection on edge devices without requiring backpropagation. Our framework leverages cloud resources for training while ensuring that privacy-sensitive data never leaves the device unprotected during inference. At its core, SecDOOD consists of a HyperNetwork-based \cite{ha2022hypernetworks} personalized parameter generation module that adapts cloud-trained models to device-specific data distributions without requiring on-device fine-tuning, ensuring that the model remains effective across diverse deployment environments. To further enhance efficiency, we introduce a dynamic feature sampling and encryption strategy that selectively encrypts the most informative feature channels before transmission to the cloud, significantly reducing encryption overhead while maintaining the effectiveness of the HyperNetwork for generating personalized parameters. By integrating these components, SecDOOD enables efficient and privacy-preserving OOD detection, making it feasible for deployment on resource-constrained edge devices.

To address these challenges, we present \textbf{SecDOOD} (\textbf{Sec}ure On-\textbf{D}evice \textbf{OOD} Detection), a cloud-device collaboration approach for secure and efficient OOD detection on edge devices that does \textit{not} require local backpropagation. 
\method leverages cloud-based training while ensuring that sensitive user data remains protected on-device during inference. 
Central to our method is a HyperNetwork-based \cite{ha2022hypernetworks} personalized parameter generation module, which tailors cloud-trained models to device-specific data distributions without on-device fine-tuning. 
To further improve efficiency, we introduce a dynamic feature sampling and encryption procedure that selectively encrypts only the most informative feature channels before transmission, minimizing encryption overhead while retaining model effectiveness. 
This design enables privacy-preserving OOD detection that is suitable for resource-constrained edge hardware.

% In summary, our contributions are as follows:
% \begin{itemize}
%     \item We propose a cloud-device collaboration framework for on-device OOD detection that eliminates the need for backpropagation on device, making deployment feasible even on resource-limited devices.
%     \item We introduce the SecDOOD, which incorporates personalized parameter generation, dynamic feature sampling, and encryption strategies to achieve efficient and privacy-preserving OOD detection.
%     \item Extensive experiments on multiple datasets and OOD detection tasks demonstrate that SecDOOD achieves performance comparable to fully fine-tuned models, validating its effectiveness in real-world deployment scenarios.
% \end{itemize}

In summary, our contributions are as follows:
\begin{itemize}
\item \textbf{Backpropagation-Free Deployment.} We propose a cloud-device collaboration way for on-device OOD detection \textit{without} local backpropagation by hypernetworks, enabling deployment on resource-limited edge devices.
\item \textbf{Personalized Privacy Preservation.} SecDOOD incorporates personalized parameter generation, dynamic feature sampling, and selective encryption to safeguard user data and maintain detection performance.
\item \textbf{Robust and Efficient Performance.} Through comprehensive evaluations on multiple datasets and OOD tasks, we demonstrate that SecDOOD achieves performance comparable to fully fine-tuned models.
% , confirming its applicability in real-world scenarios.
\end{itemize}

\section{Related Works}
\noindent \textbf{Out-of-Distribution Detection.}  
The task of Out-of-Distribution (OOD) detection involves identifying samples that deviate from the training distribution while ensuring that in-distribution (ID) classification remains accurate. Methods for OOD detection generally fall into two categories: post hoc techniques and training-time regularization \cite{yang2022openood}. Post hoc methods analyze model outputs to compute OOD scores, as seen in approaches like Maximum Softmax Probability (MSP) \cite{oodbaseline17iclr}. Enhancements to these methods include temperature scaling and input perturbation techniques \cite{Liang_2018_ECCV}, while later improvements such as MaxLogit \cite{hendrycks2019anomalyseg} and energy-based scoring \cite{energyood20nips} have refined detection accuracy. Feature-space modifications, including activation-trimming strategies like ReAct \cite{sun2021tone} and ASH \cite{djurisic2022extremely}, as well as distance-based classifiers such as Mahalanobis \cite{mahananobis18nips} and k-Nearest Neighbor (kNN) \cite{sun2022knnood}, leverage learned feature representations for OOD detection. Hybrid models have emerged, such as VIM \cite{haoqi2022vim} and Generalized Entropy (GEN) \cite{liu2023gen}, which incorporate both feature statistics and logits to improve detection robustness.

\noindent \textbf{Cloud-Device Collaborative Learning.}  
Cloud-device collaborative learning has gained traction as a means of optimizing computational efficiency while maintaining data privacy \cite{yao2021device, lv2023duet, lv2023ideal,Litomm}. A common implementation is federated learning (FL) \cite{collins2022fedavg,sabah2024model}, where distributed edge devices collaboratively train a shared global model without transmitting raw data to a central server. While FL preserves privacy, it incurs high communication costs due to frequent model synchronization, making it less suitable for large-scale, real-time applications. Alternative paradigms such as split learning \cite{lin2024split} and offloading-based computation \cite{li2024dual} distribute model processing between cloud servers and edge devices, either by delegating partial training to edge devices or by performing local inference while relying on cloud-based model updates. Additionally, knowledge distillation \cite{yao2021device} has been explored as a strategy for transferring pre-trained cloud model knowledge to lightweight device-side models, though such methods often require periodic fine-tuning, imposing computational overhead on resource-limited devices. In contrast, our approach eliminates the need for local model adaptation, training, or fine-tuning. Instead, we employ cloud-assisted data selection and domain adaptation to personalize model outputs while maintaining minimal computational demands on the device. This framework significantly enhances scalability and adaptability, making it a viable solution for large-scale, resource-constrained environments.
\section{Proposed Method}

\subsection{Problem Statement and Preliminaries}
Given a training set \( \mathbf{D} = \left\{ (x_i, y_i) \right\}_{i=1}^{n} \), where each \( x_i \in \mathcal{X} \) represents an input sample and \( y_i \in \mathcal{Y} = \{1, 2, \dots, C\} \) denotes its corresponding class label, the goal of OOD detection is to differentiate between ID and OOD samples. OOD samples exhibit \textit{semantic shifts} relative to ID samples and do not belong to any predefined class in \( \mathcal{Y} \). The model consists of a feature extractor \( g(\cdot) \), which derives feature representations, and a classifier \( h(\cdot) \), which generates predictions, producing a probability output \( \hat{p} \).

The ID data follow a marginal distribution \( P_{\text{in}} \), while OOD samples encountered during testing originate from a different marginal distribution \( P_{\text{out}} \). The objective of OOD detection is to learn a decision function \( G \) that determines whether a test sample \( x \in \mathcal{X} \) belongs to the ID distribution or is an OOD sample:

\begin{equation}
    G(x; g, h) = 
        \begin{cases} 
        0 & \text{if } x \sim \mathcal{D}_{\text{out}}, \\
        1 & \text{if } x \sim \mathcal{D}_{\text{in}}.
        \end{cases}
\end{equation}

\begin{problem}[On-Device OOD Detection]
In this work, we consider a setting where the OOD detection model is deployed on a device. The model is expected to use only the device's computational resources during the prediction phase in order to protect user data privacy. Let \( \mathcal{C}_{\text{device}} \) denote the limited computational capacity of the device. In this setting, for any test sample \( x \in \mathcal{X} \), the prediction function \( G(x; g, h) \) must be evaluated under the constraint \( \mathcal{C}(x) \leq \mathcal{C}_{\text{device}} \), where \( \mathcal{C}(x) \) represents the computational cost associated with processing \( x \).
\end{problem}

\noindent \textbf{Key Challenges in On-Device OOD Detection.}
One challenge is that the device's computational capacity, denoted by \( \mathcal{C}_{\text{device}} \), is limited, making heavy computations (e.g., backpropagation) infeasible on-device. Consequently, the training phase must occur offline, and the device performs only inference with a fixed model, potentially degrading the performance of \( G(x; g, h) \) on unseen OOD samples. These challenges imply constraints such as \( \mathcal{C}(\theta) \leq \mathcal{C}_{\text{device}} \) and \( G(x; g, h)|_{x\sim P_{\text{out}}} \leq \epsilon \), where \( \epsilon \) indicates the acceptable detection performance.

\begin{figure*}[t]
    \centering
    \includegraphics[width=0.8\linewidth]{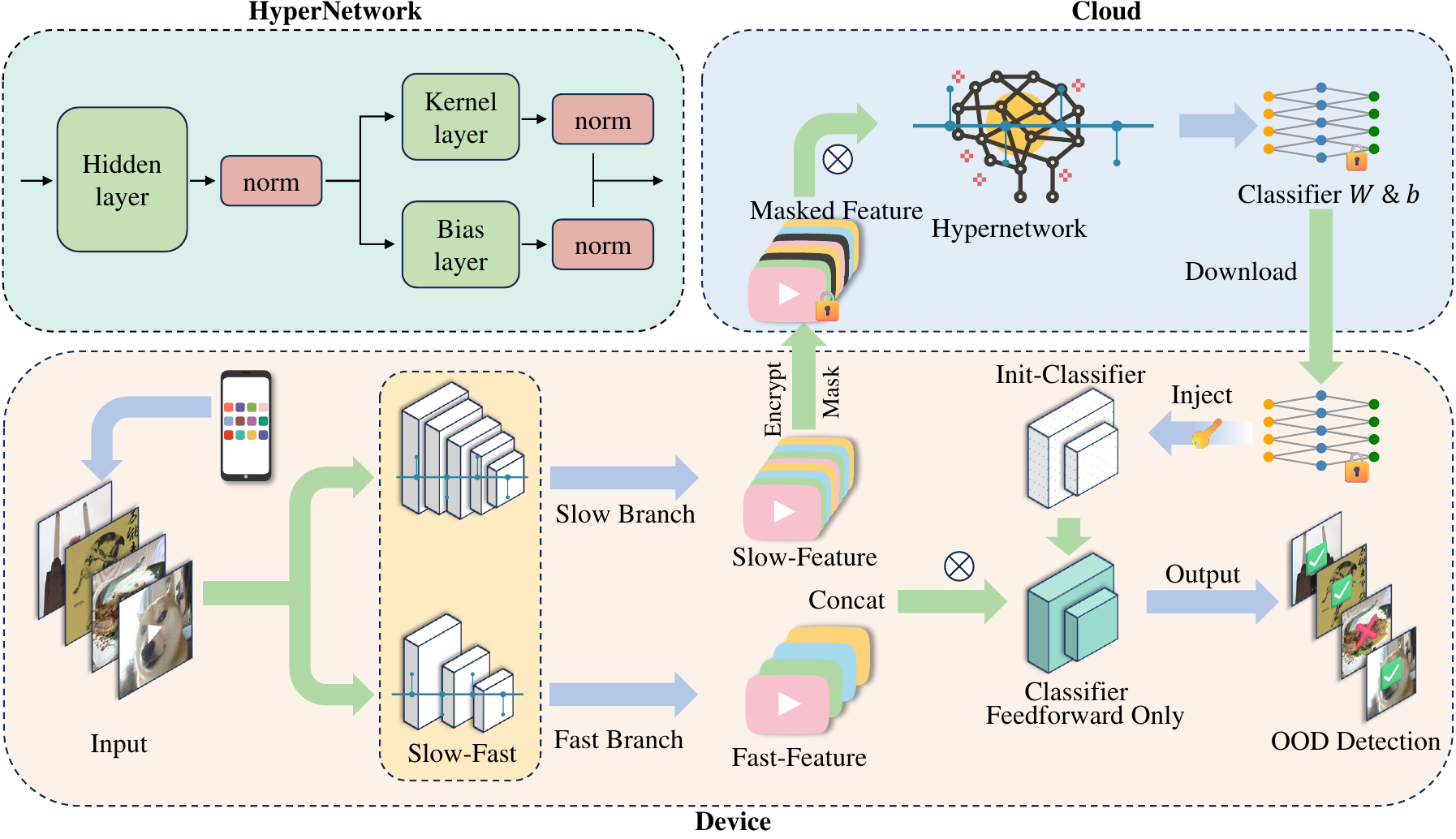} 
    \caption{Overview of the proposed \method framework. SecDOOD leverages cloud-device collaboration to enable efficient and privacy-preserving OOD detection on edge devices. During deployment, the edge device extracts feature representations from incoming data and applies dynamic feature sampling to select the most informative channels. These selected features are securely encrypted and transmitted to the cloud, where a HyperNetwork-based module generates personalized model parameters tailored to the device-specific data distribution. The personalized parameters are then sent back to the device and encrypted, where they are decrypted and used for inference. This approach ensures adaptability to user-specific distributions while maintaining computational efficiency and data privacy.}
    \label{fig:main}
    \vspace{-1em}
\end{figure*}

\subsection{Overview}
To address the problems and challenges above, it is key to design a collaboration system that leverages cloud resources for comprehensive training and processing, while preserving data privacy and meeting computational constraints. In this system, a test sample is first encrypted on the device and then uploaded to the cloud, where the encrypted real-time test samples \( En(\mathcal{S}_{T}) \) are processed. The processed results are then returned to the device for decryption. In particular, our approach trains a cloud model \( \mathcal{M}_g(\cdot; \Theta_g) \) using the personal ID samples \( \mathcal{S}_{ID} \) and transfers its knowledge to a local device model \( \mathcal{M}_c(\cdot; \Theta_c) \) to ensure that the device, operating under limited computational capacity \( \mathcal{C}_{\text{device}} \), maintains acceptable OOD detection performance.

\begin{equation}
    \textbf{SecDOOD}: 
    \underbrace{\mathcal{M}_{c}(\mathcal{S}_{ID};\Theta_c)}_{\text{Cloud Model}} 
    \rightarrow
    \underbrace{\mathcal{M}_{d}(En(\mathcal{S}_T);\Theta_{d})}_{\text{Device Model}}.
\end{equation}

\noindent \textbf{Framework Pipeline.} Our framework operates in three distinct stages. In the first stage, the cloud model \( \mathcal{M}_g(\cdot) \) is trained using the ID samples \( \mathcal{S}_{ID} \), incorporating a HyperNetwork-based module that generates personalized parameters. In the second stage, the device extracts features $ F_{T} $ from its real-time samples \( \mathcal{S}_{T} \) using a pre-trained vision backbone, which remains fixed during the process. The extracted features are then encrypted and uploaded to the cloud. The cloud model \( \mathcal{M}_g(\cdot) \) processes these encrypted features to derive personalized parameters for the device model. In the final stage, the device receives and decrypts the updated parameters from the cloud, applies them to its local model \( \mathcal{M}_{d}(\cdot) \), and produces the final predictions.

% As shown in Fig. ~\ref{}, 

\subsection{Personalized Parameters Generation}
\noindent \textbf{Motivation.} Adapting the cloud model to real-time samples on a personalized device usually requires on-device fine-tuning, which is difficult due to the computation constraint. To overcome this issue, we propose a HyperNetwork-based \cite{ha2022hypernetworks} personalized parameters generation module implemented as a cloud service. This module processes videos captured by the device and generates device-specific parameters that are designed to adapt the model to the unique data distributions present on the device.

We leverage a HyperNetwork-based approach to generate personalized parameters. Given the cloud model \( \mathcal{M}_g(\cdot; \Theta_g) \) trained on ID samples \( \mathcal{S}_{ID} \), we define a HyperNetwork \( H(\cdot; \Theta_H) \) that learns a mapping from the device-extracted features \( F_T \) to personalized model parameters \( \Theta_d \) for the device model \( \mathcal{M}_d(\cdot; \Theta_d) \). Formally, the parameter adaptation process can be written as:

\begin{equation}
    \Theta_d = H(F_T; \Theta_H).
\end{equation}

Here, \( H(\cdot; \Theta_H) \) is designed to generate device-specific parameters dynamically, ensuring the model can adapt to heterogeneous data distributions across different devices without requiring direct on-device updates. Instead of learning a single fixed set of parameters, the HyperNetwork learns to generalize across different device conditions, which aligns with the principles of meta-learning—learning how to adapt efficiently to new data distributions. 

\noindent \textbf{Meta-Learning Objective.}  
Since device data distributions vary, directly applying a globally trained model often results in performance degradation. To ensure effective adaptation, we formulate the learning of \( H(\cdot; \Theta_H) \) as a meta-learning problem \cite{hospedales2021meta}, where the objective is to optimize the HyperNetwork such that the generated parameters \( \Theta_d \) allow the device model to perform well on real-time samples. Given an ID dataset \( \mathcal{S}_{ID} \) used to train the cloud model and a device-specific real-time dataset \( \mathcal{S}_{T} \), the optimization is formulated as:

\begin{equation}
    \min_{\Theta_H} \mathbb{E}_{x \sim \mathcal{S}_{T}} \mathcal{L}(G(x; g, h_{\Theta_d})),  \text{where}  \quad \Theta_d = H(F_T; \Theta_H).
\end{equation}

Here, \( \mathcal{L} \) represents the loss function for the OOD detection decision function \( G(x; g, h_{\Theta_d}) \), ensuring that the generated parameters \( \Theta_d \) effectively adapt to the local data distribution without requiring on-device fine-tuning. This meta-learning formulation enables the HyperNetwork to generalize its parameter generation ability across diverse device environments, making it robust to unseen data variations.

\noindent \textbf{Cloud-Device Parameter Transfer.}  
Once the HyperNetwork generates personalized parameters \( \Theta_d \), they are securely transmitted to the device for model adaptation. The device model \( \mathcal{M}_d(\cdot; \Theta_d) \) applies these updated parameters to perform inference on real-time test samples. Since the device does not perform backpropagation or fine-tuning, this approach efficiently adapts the model while maintaining computational feasibility within the device’s resource constraints \( \mathcal{C}_{\text{device}} \).
By leveraging HyperNetworks with a meta-learning objective, the proposed framework efficiently personalizes OOD detection models for heterogeneous device environments. This enables the system to achieve high detection performance without requiring computationally expensive on-device updates.

However, the parameter generation process relies on features uploaded from the device, which poses a potential risk of user privacy leakage. To mitigate this, we introduce a dynamic feature sampling and encryption module that selectively encrypts critical feature components before transmission. This design ensures that user privacy is protected while preserving the functionality of the HyperNetwork for effective parameter generation.

\subsection{Dynamic Feature Sampling and Encryption}

\noindent \textbf{Motivation.}  
In order to prevent user privacy leakage, all features to be uploaded to the cloud must be encrypted on the device. However, encryption is computationally expensive due to both the inherent cost of encryption algorithms and the limited processing power of the device. Directly encrypting the entire feature tensor \( F_T \in \mathbb{R}^{C \times T \times W \times H} \) would introduce significant latency, making real-time OOD detection infeasible. To address this challenge, we propose the \textit{Dynamic Feature Sampling and Encryption} method, which selectively encrypts only the most critical feature channels while masking less informative ones. This reduces encryption time while maintaining the performance of the personalized parameter generation module.

\noindent \textbf{Channel Importance Estimation.}  
To determine the most influential channels in feature extraction, we employ a game-theoretic importance estimation method inspired by Shapley Additive Explanations \cite{antwarg2021explaining,nohara2019explanation}. Given a vision backbone \( g(\cdot) \) that extracts feature maps \( F_T \) from input samples \( x \), our goal is to quantify the contribution of each channel \( F_T^c \), where \( F_T^c \) represents the feature tensor corresponding to channel \( c \), to the overall feature representation.

The importance score \( S(c) \) of each channel \( c \) is computed using the Shapley value formulation:

\begin{equation}
    S(c) = \sum_{\mathcal{U} \subseteq \mathcal{C} \setminus \{c\}} \frac{|\mathcal{U}|! (|\mathcal{C}| - |\mathcal{U}| - 1)!}{|\mathcal{C}|!} \left[ f(\mathcal{U} \cup \{c\}) - f(\mathcal{U}) \right],
\end{equation}

where \( \mathcal{C} = \{1, 2, \dots, C\} \) represents the set of all feature channels, and \( C \) denotes the total number of channels in \( F_T \). The subset \( \mathcal{U} \) includes a selection of channels excluding \( c \), while \( f(\mathcal{U}) \) quantifies the feature importance when only channels in \( \mathcal{U} \) are considered. The term \( \frac{|\mathcal{U}|! (|\mathcal{C}| - |\mathcal{U}| - 1)!}{|\mathcal{C}|!} \) is a combinatorial coefficient that assigns a weight to each subset \( \mathcal{U} \), ensuring a fair contribution estimation for channel \( c \). Here, the factorial notation \( n! \) represents the product of all positive integers up to \( n \).

Since computing the exact Shapley values is intractable for large \( C \), we approximate them using Monte Carlo sampling. This involves iteratively estimating \( S(c) \) by randomly selecting subsets \( \mathcal{U} \) and measuring the marginal contribution of each channel. This approach enables efficient identification of the most important channels for encryption while maintaining computational feasibility.

\noindent \textbf{Dynamic Encryption and Masking Strategy.}  
To reduce encryption overhead while preserving essential information, we rank all channels based on their importance scores \( S(c) \) and selectively encrypt the top \( \alpha \) fraction, while masking the rest. Our experiments indicate that even when the HyperNetwork receives only a subset of the feature information, it maintains strong generalization capability, effectively generating personalized and parameters for the on-device model.

\begin{equation}
    \mathcal{C}_{\text{enc}} = \{ c \mid S(c) \text{ in top } \alpha C \}, \quad
    \mathcal{C}_{\text{mask}} = \mathcal{C}_{\text{all}} \setminus \mathcal{C}_{\text{enc}}.
\end{equation}

Setting \( \alpha = 0.50 \), we encrypt only the most informative 50\% of channels, denoted as:

\begin{equation}
    F_T^{\text{enc}} = \text{Encrypt}(F_T^c), \quad c \in \mathcal{C}_{\text{enc}},
\end{equation}
while the remaining 50\% are masked, leading to the following structured feature representation:

\begin{equation}
    F_T^c =
    \begin{cases} 
        \text{Encrypt}(F_T^c), & c \in \mathcal{C}_{\text{enc}}, \\
        0, & c \in \mathcal{C}_{\text{mask}}.
    \end{cases}
\end{equation}

This strategy ensures that only the most informative channels are encrypted, while the rest are masked to reduce computational overhead. We employ \textit{homomorphic encryption} \cite{acar2018survey}, which enables computation directly on encrypted data. This allows the encrypted feature subset \( F_T^{\text{enc}} \) to be transmitted to the cloud and processed by the \( H(\cdot; \Theta_H) \) without requiring decryption.
The generated parameters \( \Theta_d \) are then securely transmitted back to the device, where they are decrypted and applied for final prediction.

By encrypting only the most critical channels and leveraging homomorphic encryption, this method effectively balances privacy protection, computational efficiency, and model adaptability.
\section{Experiment}
\begin{table*}[t]
    \centering
    \renewcommand{\arraystretch}{1.2}
    \setlength{\tabcolsep}{4pt}
    \scriptsize
    \tabcolsep=0.25cm
    \begin{tabular}{l cc cc cc cc cc}
        \toprule
        \multirow{2}{*}{\textbf{Methods}} & 
        \multicolumn{2}{c}{\textbf{HMDB51}} &
        \multicolumn{2}{c}{\textbf{UCF101}} &
        \multicolumn{2}{c}{\textbf{EPIC-Kitchens}} &
        \multicolumn{2}{c}{\textbf{HAC}} &
        \multicolumn{2}{c}{\textbf{Average}} \\
        \cmidrule(lr){2-3} \cmidrule(lr){4-5} \cmidrule(lr){6-7} \cmidrule(lr){8-9} \cmidrule(lr){10-11}
        & FPR95$\downarrow$ & AUROC$\uparrow$ 
        & FPR95$\downarrow$ & AUROC$\uparrow$ 
        & FPR95$\downarrow$ & AUROC$\uparrow$ 
        & FPR95$\downarrow$ & AUROC$\uparrow$
        & FPR95$\downarrow$ & AUROC$\uparrow$ \\
        \midrule
        \multicolumn{11}{c}{\textbf{With On-Device Training}} \\
        \midrule
        MSP         & 66.83  & 75.64  & 67.32  & 71.13  & 43.37  & 86.66  & 56.17  & 79.50  & 58.42  & 78.23 \\
        Energy      & 72.64  & 71.75  & 70.12  & 71.49  & 43.66  & 82.05  & 61.50  & 74.99  & 61.98  & 75.07 \\
        MaxLogit    & 72.06  & 73.68  & 70.47  & 71.96  & 39.84  & 84.76  & 57.95  & 77.27  & 60.08  & 76.92 \\
        ReAct       & 82.17  & 67.09  & 78.44  & 67.64  & 53.49  & 78.07  & 75.11  & 70.04  & 72.80  & 70.71 \\
        ASH         & 71.62  & 76.66  & 69.36  & 72.38  & 34.38  & 88.05  & 47.85  & 83.49  & 55.80  & 80.15 \\
        GEN         & 68.47  & 78.43  & 64.80  & 73.97  & 36.81  & 85.11  & 49.53  & 83.67  & 54.90  & 80.30 \\
        KNN         & 71.08  & 78.84  & 68.62  & 74.33  & 41.83  & 82.32  & 57.00  & 82.53  & 59.63  & 79.51 \\
        VIM         & 72.25  & 71.88  & 70.72  & 70.58  & 43.14  & 82.69  & 59.48  & 75.46  & 61.40  & 75.15 \\
        LogitNorm   & 66.48  & 79.08  & 63.79  & 75.10  & 39.03  & 85.27  & 54.22  & 81.83  & 55.88  & 80.30 \\
        \midrule
        \multicolumn{11}{c}{\textbf{Without On-Device Training}} \\
        \midrule
        Ini-Classifier  & 84.89  & 66.19  & 85.43  & 62.65  & 87.18  & 57.94  & 89.03  & 63.42  & 86.63  & 62.55 \\
        Ini-Hypernetwork  & 86.01  & 67.31  & 84.61  & 63.05  & 88.86  & 57.86  & 88.71  & 56.72  & 87.05  & 61.24 \\
        Ours        & 69.52  & 79.25  & 69.34  & 75.37  & 31.37  & 89.01  & 58.32  & 84.98  & 57.14  & 82.15 \\
        \bottomrule
    \end{tabular}
    \caption{Far-OOD Detection results using Kinetics-600 as the ID dataset ($\uparrow$ higher is better; $\downarrow$ lower is better). Ini-Classifier refers to a randomly initialized classifier used directly for inference, while Ini-Hypernetwork denotes a randomly initialized hypernetwork that generates parameters for the classifier.}
    \label{tab:far_kinetics}
\end{table*}
\begin{table*}[t]
    \centering
    \renewcommand{\arraystretch}{1.2}
    \setlength{\tabcolsep}{4pt}
    \scriptsize
    \tabcolsep=0.25cm
    \begin{tabular}{l cc cc cc cc cc}
        \toprule
        \multirow{2}{*}{\textbf{Methods}} & 
        \multicolumn{2}{c}{\textbf{Kinetics-600}} &
        \multicolumn{2}{c}{\textbf{UCF101}} &
        \multicolumn{2}{c}{\textbf{EPIC-Kitchens}} &
        \multicolumn{2}{c}{\textbf{HAC}} &
        \multicolumn{2}{c}{\textbf{Average}} \\
        \cmidrule(lr){2-3} \cmidrule(lr){4-5} \cmidrule(lr){6-7} \cmidrule(lr){8-9} \cmidrule(lr){10-11}
        & FPR95$\downarrow$ & AUROC$\uparrow$ 
        & FPR95$\downarrow$ & AUROC$\uparrow$ 
        & FPR95$\downarrow$ & AUROC$\uparrow$
        & FPR95$\downarrow$ & AUROC$\uparrow$
        & FPR95$\downarrow$ & AUROC$\uparrow$ \\
        \midrule
        \multicolumn{11}{c}{\textbf{With On-Device Training}} \\
        \midrule
        MSP         & 39.11  & 88.78  & 46.64  & 86.40  & 17.33  & 95.99  & 39.91  & 89.10  & 35.75  & 90.07 \\
        Energy      & 32.95  & 92.48  & 44.93  & 87.95  & 8.10   & 97.70  & 32.95  & 92.28  & 29.73  & 92.60 \\
        MaxLogit    & 33.07  & 92.31  & 44.93  & 88.02  & 9.12   & 97.77  & 33.06  & 92.17  & 30.05  & 92.57 \\
        ReAct       & 27.59  & 93.54  & 44.01  & 88.05  & 7.53   & 97.61  & 31.01  & 92.86  & 27.54  & 93.02 \\
        ASH         & 51.20  & 87.81  & 53.93  & 84.22  & 19.95  & 95.92  & 42.99  & 90.23  & 42.02  & 89.55 \\
        GEN         & 41.51  & 90.34  & 46.18  & 87.91  & 8.21   & 98.26  & 38.31  & 91.28  & 33.55  & 91.95 \\
        KNN         & 22.69  & 95.01  & 39.34  & 89.28  & 9.92   & 97.92  & 20.75  & 96.02  & 23.18  & 95.06 \\
        VIM         & 13.68  & 97.01  & 33.87  & 91.45  & 5.93   & 98.15  & 13.45  & 97.12  & 16.73  & 93.93 \\
        LogitNorm   & 46.07  & 87.41  & 49.03  & 85.96  & 15.96  & 96.30  & 47.09  & 87.64  & 39.54  & 89.33 \\
        \midrule
        \multicolumn{11}{c}{\textbf{Without On-Device Training}} \\
        \midrule
        Ini-Classifier  & 94.18  & 43.60  & 96.58  & 40.51  & 97.49  & 41.39  & 85.40  & 59.71  & 93.41  & 46.30 \\
        Ini-Hypernetwork  & 89.97  & 48.99  & 86.77  & 53.69  & 90.36  & 57.22  & 92.13  & 57.28  & 89.81  & 54.30 \\
        \method       & 15.39  & 96.06  & 46.18  & 86.19  & 5.01   & 98.72  & 22.35  & 94.41  & 22.23  & 93.85 \\
        \bottomrule
    \end{tabular}
    \caption{Far-OOD Detection results using HMDB as the ID dataset ($\uparrow$ higher is better; $\downarrow$ lower is better). Ini-Classifier refers to a randomly initialized classifier used directly for inference, while Ini-Hypernetwork denotes a randomly initialized hypernetwork that generates parameters for the classifier.}
    \label{tab:far_hmdb}
\end{table*}

\begin{table*}[t]
    \centering
    \renewcommand{\arraystretch}{1.2}
    \setlength{\tabcolsep}{4pt}
    \scriptsize
    \tabcolsep=0.35cm
    \begin{tabular}{l cc cc cc cc}
        \toprule
        \multirow{2}{*}{\textbf{Methods}} & 
        \multicolumn{2}{c}{\textbf{HMDB51}} &
        \multicolumn{2}{c}{\textbf{UCF101}} &
        \multicolumn{2}{c}{\textbf{EPIC-Kitchens}} &
        \multicolumn{2}{c}{\textbf{Kinetics-600}} \\
        \cmidrule(lr){2-3} \cmidrule(lr){4-5} \cmidrule(lr){6-7} \cmidrule(lr){8-9}
        & FPR95$\downarrow$ & AUROC$\uparrow$ 
        & FPR95$\downarrow$ & AUROC$\uparrow$
        & FPR95$\downarrow$ & AUROC$\uparrow$
        & FPR95$\downarrow$ & AUROC$\uparrow$ \\
        \midrule
        \multicolumn{9}{c}{\textbf{With On-Device Training}} \\
        \midrule
        MSP         & 44.66  & 87.74  & 22.14  & 95.73  & 76.31  & 67.59  & 64.08  & 76.16 \\
        Energy      & 43.36  & 87.46  & 22.52  & 96.06  & 76.68  & 68.29  & 68.75  & 75.49 \\
        MaxLogit    & 43.36  & 87.75  & 22.52  & 96.02  & 76.68  & 68.29  & 68.73  & 75.98 \\
        Mahalanobis & 40.31  & 85.28  & 12.14  & 97.14  & 98.69  & 42.99  & 93.51  & 35.83 \\
        ReAct       & 42.05  & 87.79  & 25.63  & 95.85  & 83.96  & 65.89  & 72.40  & 73.80 \\
        ASH         & 53.59  & 87.16  & 32.14  & 94.02  & 76.87  & 68.52  & 69.03  & 75.33 \\
        GEN         & 43.79  & 87.49  & 23.79  & 95.54  & 75.93  & 63.60  & 69.24  & 76.16 \\
        KNN         & 42.92  & 88.06  & 15.63  & 96.93  & 77.05  & 65.60  & 68.67  & 74.64 \\
        VIM         & 36.82  & 88.06  & 12.52  & 97.66  & 80.97  & 63.41  & 68.77  & 75.47 \\
        LogitNorm   & 48.84  & 87.65  & 19.61  & 95.85  & 80.97  & 63.41  & 67.32  & 75.84 \\
        \midrule
        \multicolumn{9}{c}{\textbf{Without On-Device Training}} \\
        \midrule
        Baseline Classifier  & 94.77  & 52.27  & 80.58  & 66.13  & 87.74  & 58.30  & 89.19  & 57.50 \\
        Baseline Hypernetwork  & 95.21  & 54.04  & 81.81  & 59.69  & 96.27  & 46.66  & 91.29  & 55.17 \\
        Ours       & 37.03  & 88.66  & 7.48  & 98.39  & 62.69  & 74.93  & 62.36  & 76.92 \\
        \bottomrule
    \end{tabular}
    \caption{Comparison of Multimodal Near-OOD detection methods with and without on-device training across multiple datasets ($\uparrow$ the higher the better; $\downarrow$ the lower the better).}
    \label{tab:full_near}
\end{table*}

\subsection{Datasets and Evaluation Metrics}
\label{subsec:datasets_eva}
\noindent \textbf{Dataset.}
we evaluate our method across five datasets: HMDB51 \cite{kuehne2011hmdb}, UCF101 \cite{soomro2012ucf101}, Kinetics-600 \cite{kay2017kinetics}, HAC \cite{dong2023SimMMDG}, and EPIC-Kitchens \cite{Damen2018EPICKITCHENS}. More details on these datasets can be found in Appx.~\ref{appx:datasets}.

\noindent \textbf{Evaluation Metrics.}
We evaluate the performance via the use of the following metrics: (1) the false positive rate (FPR95, the lower, the better) of OOD samples when the true positive rate of ID samples is at 95\%, (2) the area under the receiver operating characteristic curve (AUROC).

\subsection{Tasks, Baseline Designs, and Implementation Details}
\label{subsec:task_base_imple}
\noindent \textbf{Tasks.}
We evaluate \method on two tasks: Near-OOD detection and Far-OOD detection \cite{dong2024multiood}. See details in Appx.~\ref{appx:tasks}

\noindent \textbf{Baseline Design.} 
As our work represents the first attempt to address the OOD detection problem in a real-world deployment scenario, there are no established baseline models for direct comparison. Therefore, we evaluate \method against conventional baseline methods that require fully training or fine-tuning both a classifier and a visual backbone on the device. 
Additionally, to further demonstrate the effectiveness of \method, we design two alternative baselines that, like our approach, do not involve on-device training. 
See details in Appx.~\ref{appx:baseline}.

\noindent \textbf{Implementation Details and Hardware.}
In both the Near-OOD and Far-OOD tasks, the batch size is set to 16, the optimizer used is Adam, and the learning rate is set to 0.0001.
See more details on implementation and hardware in Appx.~\ref{appx:im_details}.

\subsection{Performance and Generalization}
Tables~\ref{tab:far_kinetics}, \ref{tab:far_hmdb} and \ref{tab:full_near} consistently show the effectiveness of \method in comparison to on-device training methods on 2 tasks with all 5 datasets.
\method consistently achieves the best or near-best values for key metrics, including FPR95 and AUROC.

\noindent \textbf{\method Excels in Real-World Scenarios.}  
In real-world deployment, device users must effectively detect OOD data originating from diverse distributions. To simulate these conditions, we design Far-OOD detection experiments, where a single dataset is used as the ID data for training (representing users' in-distribution samples), while four additional datasets serve as OOD data (representing diverse unseen distributions). As presented in Tab.~\ref{tab:far_kinetics} and Tab.~\ref{tab:far_hmdb}, \method significantly outperforms other 'Without On-Device Training' baseline methods, demonstrating its effectiveness. In addition, \method achieves superior results compared to most 'With On-Device Training' methods. Notably, on the EPIC-Kitchens dataset, \method surpasses the best 'With On-Device Training' method by +5.12 in AUROC and -10.36 in FPR95 average. This performance not only highlights the effectiveness of \method in real-world OOD detection but also underscores the strength of our training strategy, which enables superior results even compared to fully trained models.

\noindent \textbf{\method Demonstrates Efficiency Across Different Tasks.}  
Beyond evaluating \method on far-OOD detection, we also examine its performance on a more challenging near-OOD scenario, where ID and OOD data originate from the same distribution. In this case, the OOD detection model must be highly effective in distinguishing ID from OOD samples due to their inherent similarity. As shown in Tab.~\ref{tab:full_near}, \method consistently outperforms traditional baseline methods that require on-device training. Notably, on the UCF101 dataset, \method reduces FPR95 by 38.4\% compared to the best 'With On-Device Training' method, highlighting its superior capability in handling near-OOD detection.

\noindent \textbf{\method is Effective Across Diverse Datasets.}  
To validate the effectiveness of \method, we conduct experiments on five datasets, each representing distinct distributions, including kitchen environments, YouTube videos, and everyday scenes. Among them, Kinetics-600 stands out as a particularly challenging dataset, comprising 480,000 video samples. Despite the diversity in data distributions, \method consistently delivers strong performance across all datasets, demonstrating its robustness and adaptability in various real-world scenarios.

\subsection{In-Depth Analysis}
\label{subsec:IDA}
To gain a deeper understanding of the efficiency and effectiveness of our method, we conduct a comprehensive analysis across multiple aspects of computational performance and model behavior. Specifically, we evaluate: (1) the effect of masking different numbers of channels on overall performance, (2) a FLOPs comparison between the baseline and our proposed model, and (3) the per-sample encryption and decryption time under varying feature channel settings. These analyses are presented in the main paper. Additionally, we examine (4) the communication latency between the device and the cloud under different network conditions, which is provided in the Appx.~\ref{appx:addi_res}.

\begin{table}[t]
    \centering
    \renewcommand{\arraystretch}{1.2}
    \setlength{\tabcolsep}{3pt}
    \scriptsize
    
    \begin{tabular}{l cc cc cc}
        \toprule
        \multirow{2}{*}{\textbf{Methods}} & 
        \multicolumn{2}{c}{\textbf{No Mask}} &
        \multicolumn{2}{c}{\textbf{Mask 50\% Channels}} &
        \multicolumn{2}{c}{\textbf{Mask 75\% Channels}}\\
        \cmidrule(lr){2-3} \cmidrule(lr){4-5} \cmidrule(lr){6-7}
         & FPR95$\downarrow$ & AUROC$\uparrow$ 
         & FPR95$\downarrow$ & AUROC$\uparrow$ 
         & FPR95$\downarrow$ & AUROC$\uparrow$\\
        \midrule
        HMDB      & 37.69  & 88.72  & 37.03  & 88.66  & 45.75   & 85.86\\
        UCF      & 11.55  & 95.81  & 7.48  & 98.36  & 13.69  & 96.28\\
        EPIC       & 63.43  & 75.36  & 62.69  & 74.93  & 63.43   & 71.56\\
        Kinetics  & 62.63  & 76.36  & 62.36  & 76.92  & 63.85   & 76.56\\
        \bottomrule
    \end{tabular}
    \caption{Near-OOD Detection results using various mask proportion ($\uparrow$ higher is better; $\downarrow$ lower is better).}
    \label{tab:mask_near}
\end{table}
\begin{table}[h]
    \centering
    \renewcommand{\arraystretch}{1.2}
    \setlength{\tabcolsep}{3pt}
    \scriptsize
    \begin{tabular}{l cc cc cc}
        \toprule
        \multirow{2}{*}{\textbf{Methods}} & 
        \multicolumn{2}{c}{\textbf{No Mask}} &
        \multicolumn{2}{c}{\textbf{Randomly Mask 50\%}} &
        \multicolumn{2}{c}{\textbf{Dynamic Mask 50\%}}\\
        \cmidrule(lr){2-3} \cmidrule(lr){4-5} \cmidrule(lr){6-7}
         & FPR95$\downarrow$ & AUROC$\uparrow$ 
         & FPR95$\downarrow$ & AUROC$\uparrow$ 
         & FPR95$\downarrow$ & AUROC$\uparrow$\\
        \midrule
        \multicolumn{7}{c}{\textbf{Near-OOD}} \\ 
        \midrule
        HMDB      & 37.69  & 88.72  & 39.22  & 89.31  & 37.03   & 88.66\\
        UCF      & 11.55  & 95.81  & 12.62 & 96.32  & 7.48  & 98.36\\
        EPIC  & 63.43  & 75.36  & 63.42  & 71.69  & 62.69  & 74.93\\
        Kinetics  & 62.63  & 76.36  & 64.22  & 76.05  & 62.36 & 76.92\\
        \midrule
        \multicolumn{7}{c}{\textbf{Far-OOD (HMDB as ID)}} \\ 
        \midrule
        Kinetics      & 21.89  & 94.29  & 16.89  & 95.90  & 15.39   & 96.06\\
        UCF      & 52.79  & 81.93  & 47.54  & 86.24  & 46.18  & 86.19\\
        HAC  & 29.42  & 94.02  & 29.65  & 93.70  & 22.35  & 94.41\\
        \midrule
        \multicolumn{7}{c}{\textbf{Far-OOD (Kinetics as ID)}} \\ 
        \midrule
        HMDB      & 69.67  & 78.09  & 69.84  & 77.36  & 69.52   & 79.25\\
        UCF      & 69.63  & 72.98  & 70.89  & 72.84  & 69.34  & 75.37\\
        HAC  & 68.01  & 76.81  & 69.15  & 75.38  & 58.32   & 84.98\\
        \bottomrule
    \end{tabular}
    \caption{OOD Detection results using various mask strategies ($\uparrow$ higher is better; $\downarrow$ lower is better).}
    \label{tab:mask_method}
    \vspace{-1em}
\end{table}

\noindent \textbf{Number of Channels Masked.}  
To evaluate the effectiveness of our dynamic feature sampling approach, we conduct experiments on \method with different feature masking percentages: (1) masking 75\% of the channels, (2) masking 50\%, and (3) no masking applied, as shown in Tab.~\ref{tab:mask_near} and Appx.~\ref{tab:mask_far}. Additionally, we introduce a random masking strategy as a baseline to further highlight the advantages of our dynamic masking mechanism. As shown in Tab.~\ref{tab:mask_method}, an interesting observation is that masking 50\% of the channels often yields comparable or even better performance than applying no masking at all.  This suggests that our dynamic masking method effectively filters out noisy and less informative features, allowing the model to focus on more relevant representations. Furthermore, the results indicate that the HyperNetwork is highly robust to variations in input features, adapting effectively to the refined feature space and maintaining strong detection performance.

\begin{figure}[t]
    \centering
    \includegraphics[width=0.95\linewidth]{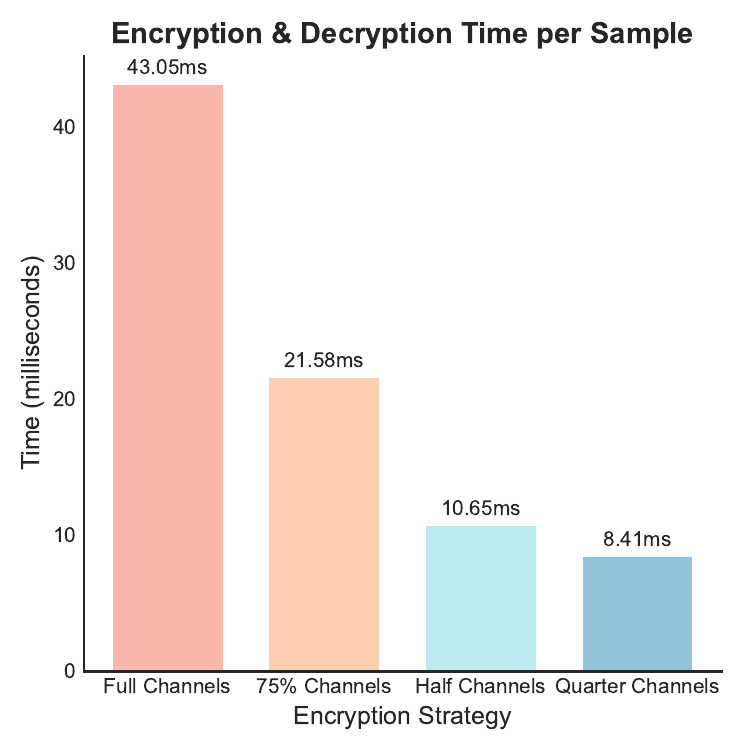} 
    \caption{Encryption and decryption time per sample under different feature channel encryption ratios. Encrypting all channels (100\%) results in significantly higher time costs, while encrypting 50\% and 25\% of the channels shows similar computational efficiency. Based on this analysis, we select 50\% encryption as the optimal balance between security and efficiency.}
    \label{fig:enc&dec}
\end{figure}

\noindent \textbf{Encryption \& Decryption Time with Different Feature Channels.}  
We evaluate the encryption and decryption time for a single sample under different feature channel encryption ratios. Specifically, we consider four settings: encrypting 25\%, 50\%, 75\%, and 100\% of the channels. As shown in Fig.~\ref{fig:enc&dec}, encrypting all channels (100\%) incurs a significantly higher time cost compared to encrypting 50\% or 25\% of the channels. Notably, the time required for encrypting \& decryption 50\% and 25\% of the channels is relatively similar, suggesting diminishing returns in efficiency gains for lower encryption ratios. Considering both computational efficiency and the impact of different encryption ratios on model performance, we ultimately select the 50\% encryption setting as the optimal balance.

\begin{figure}[t]
    \centering
    \includegraphics[width=0.95\linewidth]{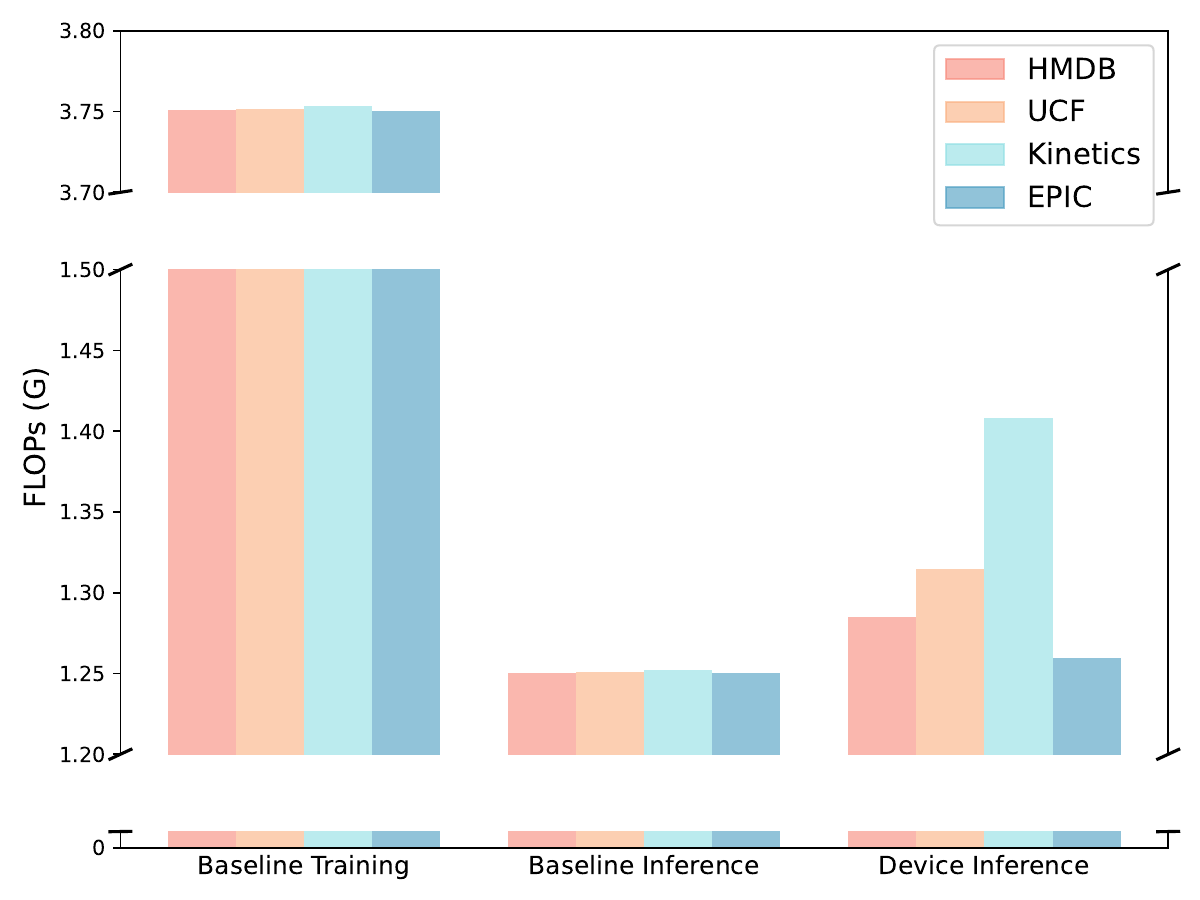} 
    \caption{FLOPs comparison between traditional on-device training and \method. Traditional methods require full model training on the device, incurring high computational costs. \method offloads training to the cloud, reducing on-device computation to feature extraction and inference, achieving a 3× efficiency gain.}
    \label{fig:flops}
    \vspace{-1.7em}
\end{figure}

\noindent \textbf{FLOPs: Traditional On-Device Training vs. \method.}
Traditional on-device training methods require both training a classifier and fine-tuning the visual backbone, resulting in substantial computational overhead. In contrast, \method offloads the training process to the cloud, eliminating the need for on-device training. Specifically, our approach trains a HyperNetwork in the cloud, which generates personalized classifier parameters while keeping the visual backbone fixed. During deployment, the edge device only performs feature extraction and classifier inference (forward propagation), significantly reducing computational demands.
To quantify this efficiency gain, we compare the FLOPs of the traditional training paradigm with \method. As shown in Tab.~\ref{fig:flops}, \method reduces computational costs by a factor of three compared to traditional methods that require full on-device training. This reduction underscores the advantage of our cloud-assisted approach, enabling efficient and scalable OOD detection on resource-constrained edge devices. However, when comparing the FLOPs of \method to baseline inference, we observe a modest increase of 16\%, primarily due to encryption and decryption operations on the device.

\section{Conclusion, Limitations, and Future Work}
In this paper, we introduced SecDOOD, a secure and efficient cloud-device collaboration framework for out-of-distribution detection on resource-constrained devices. SecDOOD eliminates the need for backpropagation by leveraging a HyperNetwork-based personalized parameter generation module, enabling effective adaptation to user-specific data distributions. Additionally, the proposed dynamic feature sampling and encryption strategy reduces encryption overhead while preserving model performance, ensuring privacy and efficiency during cloud-device interactions.

\noindent \textbf{Limitations and Future Works.} Despite its advantages, SecDOOD has certain limitations. It assumes stable cloud connectivity, which may not always be available in real-world scenarios. Future work will focus on reducing the cost of cloud-device communication, and optimizing the framework for ultra-low-power devices, further broadening its applicability in diverse real-world environments.

\noindent \textbf{Broader Impact and Ethics Statement}

\noindent \textbf{Broader Impact Statement:}
SecDOOD advances on-device OOD detection by addressing the challenges of user-specific data distributions, which is critical for applications such as healthcare, autonomous driving, and smart surveillance. By enabling efficient and privacy-preserving OOD detection on resource-constrained edge devices, SecDOOD ensures reliable model performance in dynamic environments where real-time adaptation and data privacy are essential. This framework enhances model robustness, allowing systems to remain effective when encountering novel or evolving data patterns across diverse deployment scenarios.

\noindent \textbf{Ethics Statement:}
Our study emphasizes the importance of privacy, fairness, and transparency. SecDOOD safeguards user data by ensuring that unencrypted information and its extracted features remain on-device, significantly reducing privacy risks. Its personalized parameter generation effectively mitigates bias and promotes fair and reliable OOD detection while minimizing unintended behaviors of model.
{
    \small
    \bibliographystyle{ieeenat_fullname}
    \bibliography{main}

\begin{thebibliography}{49}
\providecommand{\natexlab}[1]{#1}
\providecommand{\url}[1]{\texttt{#1}}
\expandafter\ifx\csname urlstyle\endcsname\relax
  \providecommand{\doi}[1]{doi: #1}\else
  \providecommand{\doi}{doi: \begingroup \urlstyle{rm}\Url}\fi

\bibitem[Acar et~al.(2018)Acar, Aksu, Uluagac, and Conti]{acar2018survey}
Abbas Acar, Hidayet Aksu, A~Selcuk Uluagac, and Mauro Conti.
\newblock A survey on homomorphic encryption schemes: Theory and implementation.
\newblock \emph{ACM Computing Surveys (Csur)}, 51\penalty0 (4):\penalty0 1--35, 2018.

\bibitem[Antwarg et~al.(2021)Antwarg, Miller, Shapira, and Rokach]{antwarg2021explaining}
Liat Antwarg, Ronnie~Mindlin Miller, Bracha Shapira, and Lior Rokach.
\newblock Explaining anomalies detected by autoencoders using shapley additive explanations.
\newblock \emph{Expert systems with applications}, 186:\penalty0 115736, 2021.

\bibitem[Bai et~al.(2024)Bai, Han, Cao, Jiang, Hu, and Zhang]{Bai_2024_CVPR}
Yichen Bai, Zongbo Han, Bing Cao, Xiaoheng Jiang, Qinghua Hu, and Changqing Zhang.
\newblock Id-like prompt learning for few-shot out-of-distribution detection.
\newblock In \emph{CVPR}, pages 17480--17489, 2024.

\bibitem[Cho et~al.(2023)Cho, Park, and Choo]{cho2023training}
Wonwoo Cho, Jeonghoon Park, and Jaegul Choo.
\newblock Training auxiliary prototypical classifiers for explainable anomaly detection in medical image segmentation.
\newblock In \emph{WACV}, pages 2624--2633, 2023.

\bibitem[Collins et~al.(2022)Collins, Hassani, Mokhtari, and Shakkottai]{collins2022fedavg}
Liam Collins, Hamed Hassani, Aryan Mokhtari, and Sanjay Shakkottai.
\newblock Fedavg with fine tuning: Local updates lead to representation learning.
\newblock \emph{NeurIPS}, 2022.

\bibitem[Damen et~al.(2018)Damen, Doughty, Farinella, Fidler, Furnari, Kazakos, Moltisanti, Munro, Perrett, Price, and Wray]{Damen2018EPICKITCHENS}
Dima Damen, Hazel Doughty, Giovanni~Maria Farinella, Sanja Fidler, Antonino Furnari, Evangelos Kazakos, Davide Moltisanti, Jonathan Munro, Toby Perrett, Will Price, and Michael Wray.
\newblock Scaling egocentric vision: The epic-kitchens dataset.
\newblock In \emph{ECCV}, 2018.

\bibitem[Djurisic et~al.(2022)Djurisic, Bozanic, Ashok, and Liu]{djurisic2022extremely}
Andrija Djurisic, Nebojsa Bozanic, Arjun Ashok, and Rosanne Liu.
\newblock Extremely simple activation shaping for out-of-distribution detection.
\newblock \emph{arXiv preprint arXiv:2209.09858}, 2022.

\bibitem[Dong et~al.(2022)Dong, Zhang, Xu, Ai, Gu, Lu, Kannala, and Chen]{SuperFusion}
Hao Dong, Xianjing Zhang, Jintao Xu, Rui Ai, Weihao Gu, Huimin Lu, Juho Kannala, and Xieyuanli Chen.
\newblock Superfusion: Multilevel lidar-camera fusion for long-range hd map generation.
\newblock \emph{arXiv preprint arXiv:2211.15656}, 2022.

\bibitem[Dong et~al.(2023)Dong, Nejjar, Sun, Chatzi, and Fink]{dong2023SimMMDG}
Hao Dong, Ismail Nejjar, Han Sun, Eleni Chatzi, and Olga Fink.
\newblock Sim{MMDG}: A simple and effective framework for multi-modal domain generalization.
\newblock In \emph{NeurIPS}, 2023.

\bibitem[Dong et~al.(2024)Dong, Zhao, Chatzi, and Fink]{dong2024multiood}
Hao Dong, Yue Zhao, Eleni Chatzi, and Olga Fink.
\newblock Multiood: Scaling out-of-distribution detection for multiple modalities.
\newblock In \emph{NeurIPS}, 2024.

\bibitem[Du et~al.(2022)Du, Wang, Gozum, and Li]{Du_2022_CVPR}
Xuefeng Du, Xin Wang, Gabriel Gozum, and Yixuan Li.
\newblock Unknown-aware object detection: Learning what you don't know from videos in the wild.
\newblock In \emph{CVPR}, pages 13678--13688, 2022.

\bibitem[Graham et~al.(2023)Graham, Pinaya, Tudosiu, Nachev, Ourselin, and Cardoso]{grahamdenoising}
Mark~S. Graham, Walter H.~L. Pinaya, Petru-Daniel Tudosiu, Parashkev Nachev, Sebastien Ourselin, and M.~Jorge Cardoso.
\newblock Denoising diffusion models for out-of-distribution detection.
\newblock In \emph{CVPRW}, pages 2948--2957, 2023.

\bibitem[Ha et~al.(2022)Ha, Dai, and Le]{ha2022hypernetworks}
David Ha, Andrew~M Dai, and Quoc~V Le.
\newblock Hypernetworks.
\newblock In \emph{International Conference on Learning Representations}, 2022.

\bibitem[Hao et~al.(2024)Hao, Li, Liu, Liu, Lu, Xu, Li, Chen, Yue, Fu, et~al.]{hao2024artificial}
Nan Hao, Yuangang Li, Kecheng Liu, Songtao Liu, Yingzhou Lu, Bohao Xu, Chenhao Li, Jintai Chen, Ling Yue, Tianfan Fu, et~al.
\newblock Artificial intelligence-aided digital twin design: A systematic review.
\newblock 2024.

\bibitem[Hendrycks and Gimpel(2017)]{oodbaseline17iclr}
Dan Hendrycks and Kevin Gimpel.
\newblock A baseline for detecting misclassified and out-of-distribution examples in neural networks.
\newblock In \emph{ICLR}, 2017.

\bibitem[Hendrycks et~al.(2022)Hendrycks, Basart, Mazeika, Zou, Kwon, Mostajabi, Steinhardt, and Song]{hendrycks2019anomalyseg}
Dan Hendrycks, Steven Basart, Mantas Mazeika, Andy Zou, Joe Kwon, Mohammadreza Mostajabi, Jacob Steinhardt, and Dawn Song.
\newblock Scaling out-of-distribution detection for real-world settings.
\newblock \emph{ICML}, 2022.

\bibitem[Hospedales et~al.(2021)Hospedales, Antoniou, Micaelli, and Storkey]{hospedales2021meta}
Timothy Hospedales, Antreas Antoniou, Paul Micaelli, and Amos Storkey.
\newblock Meta-learning in neural networks: A survey.
\newblock \emph{IEEE transactions on pattern analysis and machine intelligence}, 44\penalty0 (9):\penalty0 5149--5169, 2021.

\bibitem[Ji et~al.(2025)Ji, Li, Lv, Zhang, Li, Wan, Lei, and Zimmermann]{Litomm}
Wei Ji, Li Li, Zheqi Lv, Wenqiao Zhang, Mengze Li, Zhen Wan, Wenqiang Lei, and Roger Zimmermann.
\newblock Backpropagation-free multi-modal on-device model adaptation via cloud-device collaboration.
\newblock \emph{ACM Trans. Multimedia Comput. Commun. Appl.}, 21\penalty0 (2), 2025.

\bibitem[Karimi and Gholipour(2022)]{karimi2022improving}
Davood Karimi and Ali Gholipour.
\newblock Improving calibration and out-of-distribution detection in deep models for medical image segmentation.
\newblock \emph{IEEE transactions on artificial intelligence}, 4\penalty0 (2):\penalty0 383--397, 2022.

\bibitem[Kay et~al.(2017)Kay, Carreira, Simonyan, Zhang, Hillier, Vijayanarasimhan, Viola, Green, Back, Natsev, et~al.]{kay2017kinetics}
Will Kay, Joao Carreira, Karen Simonyan, Brian Zhang, Chloe Hillier, Sudheendra Vijayanarasimhan, Fabio Viola, Tim Green, Trevor Back, Paul Natsev, et~al.
\newblock The kinetics human action video dataset.
\newblock \emph{arXiv preprint arXiv:1705.06950}, 2017.

\bibitem[Kuehne et~al.(2011)Kuehne, Jhuang, Garrote, Poggio, and Serre]{kuehne2011hmdb}
Hildegard Kuehne, Hueihan Jhuang, Est{\'\i}baliz Garrote, Tomaso Poggio, and Thomas Serre.
\newblock Hmdb: a large video database for human motion recognition.
\newblock In \emph{ICCV}, 2011.

\bibitem[Lee et~al.(2018)Lee, Lee, Lee, and Shin]{mahananobis18nips}
Kimin Lee, Kibok Lee, Honglak Lee, and Jinwoo Shin.
\newblock A simple unified framework for detecting out-of-distribution samples and adversarial attacks.
\newblock In \emph{NeurIPS}, 2018.

\bibitem[Li et~al.(2024{\natexlab{a}})Li, Sun, Peng, Cheng, Yin, Cheng, Liu, Li, and Xu]{li2024dual}
Dongbo Li, Yuchen Sun, Jielun Peng, Siyao Cheng, Zhisheng Yin, Nan Cheng, Jie Liu, Zhijun Li, and Chenren Xu.
\newblock Dual network computation offloading based on drl for satellite-terrestrial integrated networks.
\newblock \emph{IEEE Transactions on Mobile Computing}, 2024{\natexlab{a}}.

\bibitem[Li et~al.(2023{\natexlab{a}})Li, Chen, He, Yu, Liu, and Jia]{Li_2023_CVPR}
Jingyao Li, Pengguang Chen, Zexin He, Shaozuo Yu, Shu Liu, and Jiaya Jia.
\newblock Rethinking out-of-distribution (ood) detection: Masked image modeling is all you need.
\newblock In \emph{CVPR}, pages 11578--11589, 2023{\natexlab{a}}.

\bibitem[Li et~al.(2024{\natexlab{b}})Li, Li, Tu, Liu, Guo, Juefei-Xu, Xu, and Yu]{li2024light}
Jinlong Li, Baolu Li, Zhengzhong Tu, Xinyu Liu, Qing Guo, Felix Juefei-Xu, Runsheng Xu, and Hongkai Yu.
\newblock Light the night: A multi-condition diffusion framework for unpaired low-light enhancement in autonomous driving.
\newblock In \emph{CVPR}, pages 15205--15215, 2024{\natexlab{b}}.

\bibitem[Li et~al.(2023{\natexlab{b}})Li, Wang, Qin, Ji, and Liang]{li2023biased}
Li Li, Chenwei Wang, You Qin, Wei Ji, and Renjie Liang.
\newblock Biased-predicate annotation identification via unbiased visual predicate representation.
\newblock In \emph{ACM MM}, pages 4410--4420, 2023{\natexlab{b}}.

\bibitem[Li et~al.(2024{\natexlab{c}})Li, Ji, Wu, Li, Qin, Wei, and Zimmermann]{li2024panoptic}
Li Li, Wei Ji, Yiming Wu, Mengze Li, You Qin, Lina Wei, and Roger Zimmermann.
\newblock Panoptic scene graph generation with semantics-prototype learning.
\newblock In \emph{AAAI}, pages 3145--3153, 2024{\natexlab{c}}.

\bibitem[Li et~al.(2024{\natexlab{d}})Li, Gong, Dong, Yang, Tu, and Zhao]{li2024dpudynamicprototypeupdating}
Shawn Li, Huixian Gong, Hao Dong, Tiankai Yang, Zhengzhong Tu, and Yue Zhao.
\newblock Dpu: Dynamic prototype updating for multimodal out-of-distribution detection, 2024{\natexlab{d}}.

\bibitem[Liang et~al.(2018)Liang, Yang, Wang, and Urtasun]{Liang_2018_ECCV}
Ming Liang, Bin Yang, Shenlong Wang, and Raquel Urtasun.
\newblock Deep continuous fusion for multi-sensor 3d object detection.
\newblock In \emph{ECCV}, 2018.

\bibitem[Lin et~al.(2024)Lin, Qu, Chen, and Huang]{lin2024split}
Zheng Lin, Guanqiao Qu, Xianhao Chen, and Kaibin Huang.
\newblock Split learning in 6g edge networks.
\newblock \emph{IEEE Wireless Communications}, 2024.

\bibitem[Liu et~al.(2020)Liu, Wang, Owens, and Li]{energyood20nips}
Weitang Liu, Xiaoyun Wang, John~D Owens, and Yixuan Li.
\newblock Energy-based out-of-distribution detection.
\newblock In \emph{NeurIPS}, 2020.

\bibitem[Liu et~al.(2023)Liu, Lochman, and Zach]{liu2023gen}
Xixi Liu, Yaroslava Lochman, and Christopher Zach.
\newblock Gen: Pushing the limits of softmax-based out-of-distribution detection.
\newblock In \emph{CVPR}, 2023.

\bibitem[Lv et~al.(2023{\natexlab{a}})Lv, Chen, Zhang, Kuang, Zhang, Li, Ooi, and Wu]{lv2023ideal}
Zheqi Lv, Zhengyu Chen, Shengyu Zhang, Kun Kuang, Wenqiao Zhang, Mengze Li, Beng~Chin Ooi, and Fei Wu.
\newblock Ideal: Toward high-efficiency device-cloud collaborative and dynamic recommendation system.
\newblock \emph{arXiv preprint arXiv:2302.07335}, 2023{\natexlab{a}}.

\bibitem[Lv et~al.(2023{\natexlab{b}})Lv, Zhang, Zhang, Kuang, Wang, Wang, Chen, Shen, Yang, Ooi, et~al.]{lv2023duet}
Zheqi Lv, Wenqiao Zhang, Shengyu Zhang, Kun Kuang, Feng Wang, Yongwei Wang, Zhengyu Chen, Tao Shen, Hongxia Yang, Beng~Chin Ooi, et~al.
\newblock Duet: A tuning-free device-cloud collaborative parameters generation framework for efficient device model generalization.
\newblock In \emph{WWW}, pages 3077--3085, 2023{\natexlab{b}}.

\bibitem[Munro and Damen(2020)]{munro20multi}
Jonathan Munro and Dima Damen.
\newblock {M}ulti-modal {D}omain {A}daptation for {F}ine-grained {A}ction {R}ecognition.
\newblock In \emph{CVPR}, 2020.

\bibitem[Nohara et~al.(2019)Nohara, Matsumoto, Soejima, and Nakashima]{nohara2019explanation}
Yasunobu Nohara, Koutarou Matsumoto, Hidehisa Soejima, and Naoki Nakashima.
\newblock Explanation of machine learning models using improved shapley additive explanation.
\newblock In \emph{Proceedings of the 10th ACM international conference on bioinformatics, computational biology and health informatics}, pages 546--546, 2019.

\bibitem[Qin et~al.(2024)Qin, Zhang, Nian, Ding, and Zhao]{qin2024metaood}
Yuehan Qin, Yichi Zhang, Yi Nian, Xueying Ding, and Yue Zhao.
\newblock Metaood: Automatic selection of ood detection models.
\newblock \emph{arXiv preprint arXiv:2410.03074}, 2024.

\bibitem[Sabah et~al.(2024)Sabah, Chen, Yang, Azam, Ahmad, and Sarwar]{sabah2024model}
Fahad Sabah, Yuwen Chen, Zhen Yang, Muhammad Azam, Nadeem Ahmad, and Raheem Sarwar.
\newblock Model optimization techniques in personalized federated learning: A survey.
\newblock \emph{Expert Systems with Applications}, 243:\penalty0 122874, 2024.

\bibitem[Soomro et~al.(2012)Soomro, Zamir, and Shah]{soomro2012ucf101}
Khurram Soomro, Amir~Roshan Zamir, and Mubarak Shah.
\newblock Ucf101: A dataset of 101 human actions classes from videos in the wild.
\newblock \emph{arXiv preprint arXiv:1212.0402}, 2012.

\bibitem[Sun et~al.(2021)Sun, Guo, and Li]{sun2021tone}
Yiyou Sun, Chuan Guo, and Yixuan Li.
\newblock React: Out-of-distribution detection with rectified activations.
\newblock In \emph{NeurIPS}, 2021.

\bibitem[Sun et~al.(2022{\natexlab{a}})Sun, Ming, Zhu, and Li]{sun2022knnood}
Yiyou Sun, Yifei Ming, Xiaojin Zhu, and Yixuan Li.
\newblock Out-of-distribution detection with deep nearest neighbors.
\newblock \emph{ICML}, 2022{\natexlab{a}}.

\bibitem[Sun et~al.(2022{\natexlab{b}})Sun, Li, Liu, Du, and Li]{sun2022icse}
Zhensu Sun, Li Li, Yan Liu, Xiaoning Du, and Li Li.
\newblock On the importance of building high-quality training datasets for neural code search.
\newblock In \emph{ICSE}, page 1609–1620, 2022{\natexlab{b}}.

\bibitem[Wang et~al.(2022{\natexlab{a}})Wang, Li, Feng, and Zhang]{haoqi2022vim}
Haoqi Wang, Zhizhong Li, Litong Feng, and Wayne Zhang.
\newblock Vim: Out-of-distribution with virtual-logit matching.
\newblock In \emph{CVPR}, 2022{\natexlab{a}}.

\bibitem[Wang et~al.(2022{\natexlab{b}})Wang, Yang, Men, Lin, Bai, Li, Ma, Zhou, Zhou, and Yang]{wang2022ofa}
Peng Wang, An Yang, Rui Men, Junyang Lin, Shuai Bai, Zhikang Li, Jianxin Ma, Chang Zhou, Jingren Zhou, and Hongxia Yang.
\newblock Ofa: Unifying architectures, tasks, and modalities through a simple sequence-to-sequence learning framework.
\newblock In \emph{ICML}, 2022{\natexlab{b}}.

\bibitem[Xu et~al.(2024)Xu, Liu, Yao, Yu, Ding, and Zhao]{xu2024lego}
Haoyan Xu, Kay Liu, Zhengtao Yao, Philip~S Yu, Kaize Ding, and Yue Zhao.
\newblock Lego-learn: Label-efficient graph open-set learning.
\newblock \emph{arXiv preprint arXiv:2410.16386}, 2024.

\bibitem[Yang et~al.(2022)Yang, Wang, Zou, Zhou, Ding, Peng, Wang, Chen, Li, Sun, et~al.]{yang2022openood}
Jingkang Yang, Pengyun Wang, Dejian Zou, Zitang Zhou, Kunyuan Ding, Wenxuan Peng, Haoqi Wang, Guangyao Chen, Bo Li, Yiyou Sun, et~al.
\newblock Openood: Benchmarking generalized out-of-distribution detection.
\newblock In \emph{NeurIPS}, 2022.

\bibitem[Yao et~al.(2021)Yao, Wang, Jia, Han, Zhou, and Yang]{yao2021device}
Jiangchao Yao, Feng Wang, Kunyang Jia, Bo Han, Jingren Zhou, and Hongxia Yang.
\newblock Device-cloud collaborative learning for recommendation.
\newblock In \emph{Proceedings of the 27th ACM SIGKDD Conference on Knowledge Discovery \& Data Mining}, pages 3865--3874, 2021.

\bibitem[Yi et~al.(2023)Yi, Chan, Lee, Boles, and Zhang]{yi2023uncertainty}
Wenting Yi, Wai~Kit Chan, Hiu~Hung Lee, Steven~T Boles, and Xiaoge Zhang.
\newblock An uncertainty-aware deep learning model for reliable detection of steel wire rope defects.
\newblock \emph{IEEE Transactions on Reliability}, 2023.

\bibitem[Zach et~al.(2007)Zach, Pock, and Bischof]{zach2007duality}
Christopher Zach, Thomas Pock, and Horst Bischof.
\newblock A duality based approach for realtime tv-l 1 optical flow.
\newblock In \emph{Pattern Recognition}, 2007.

\end{thebibliography}
}
\appendix
\newpage
%\setcounter{page}{1}
%\maketitlesupplementary

\setcounter{table}{0}
\setcounter{figure}{0}

\renewcommand{\thetable}{\Alph{table}}
\renewcommand{\thefigure}{\Alph{figure}}

\section{Additional Experimental Settings and Results}
\subsection{Datasets}
\label{appx:datasets}
As briefly discussed in \S \ref{subsec:datasets_eva}, we evaluate our method across five datasets: HMDB51 \cite{kuehne2011hmdb}, UCF101 \cite{soomro2012ucf101}, Kinetics-600 \cite{kay2017kinetics}, HAC \cite{dong2023SimMMDG}, and EPIC-Kitchens \cite{Damen2018EPICKITCHENS}.

\noindent
1) \textbf{HMDB51} \cite{kuehne2011hmdb} is a video action recognition dataset containing 6,766 video clips across 51 action categories. The clips are sourced from various media, including digitized movies and YouTube videos, and include both video and optical flow modalities.

\noindent
2) \textbf{UCF101} \cite{soomro2012ucf101} is a diverse video action recognition dataset collected from YouTube, containing 13,320 clips representing 101 actions. This dataset includes variations in camera motion, object appearance, scale, pose, viewpoint, and background conditions. It provides video and optical flow modalities.

\noindent
3) \textbf{Kinetics-600} \cite{kay2017kinetics} is a large-scale action recognition dataset with approximately 480,000 video clips across 600 action categories. Each clip is a 10-second snippet of an annotated action moment sourced from YouTube. Following \cite{dong2024multiood}, we selected a subset of 229 classes from Kinetics-600 to avoid potential overlaps with other datasets, resulting in 57,205 video clips. Video and audio modalities are available, with optical flow extracted at 24 frames per second using the TV-L1 algorithm \cite{zach2007duality}, yielding 114,410 optical flow samples.

\noindent
4) \textbf{HAC} \cite{dong2023SimMMDG} includes seven actions—such as `sleeping', `watching TV', `eating', and `running'—performed by humans, animals, and cartoon characters, with 3,381 total video clips. The dataset provides video, optical flow, and audio modalities.

\noindent
5) \textbf{EPIC-Kitchens} \cite{Damen2018EPICKITCHENS} is a large-scale egocentric video dataset collected from 32 participants in their kitchens as they captured routine activities. For our experiments, we use a subset from the Multimodal Domain Adaptation paper \cite{munro20multi}, which contains 4,871 video clips across the eight most common actions in participant P22's sequence (`put,' `take,' `open,' `close,' `wash,' `cut,' `mix,' and `pour'). The available modalities include video, optical flow, and audio.

\subsection{Tasks}
\label{appx:tasks}
As briefly discussed in \S \ref{subsec:task_base_imple}, we evaluate our method on two tasks: Near-OOD detection, and Far-OOD detection \cite{dong2024multiood}.

For Near-OOD detection, we evaluate using four datasets. In EPIC-Kitchens 4/4, a subset of the EPIC-Kitchens dataset is divided into four classes for training as ID and four classes for testing as OOD, totaling 4,871 video clips. HMDB51 25/26 and UCF101 50/51 are similarly derived from HMDB51 \cite{kuehne2011hmdb} and UCF101 \cite{soomro2012ucf101}, containing 6,766 and 13,320 video clips, respectively. For Kinetics-600 129/100, a subset of 229 classes is selected from Kinetics-600 \cite{kay2017kinetics}, with approximately 250 clips per class, totaling 57,205 clips. In this setup, 129 classes are used for training (ID) and the remaining 100 for testing (OOD). We present the results of \{video, optical flow\} on all four datasets, and the results of \{video, optical flow, audio\} on Kinetics-600 dataset.

In the Far-OOD detection setup, either HMDB51 or Kinetics-600 is used as the ID dataset, with the other datasets serving as OOD datasets:

\noindent
\textbf{HMDB51 as ID}: We designate UCF101, EPIC-Kitchens, HAC, and Kinetics-600 as OOD datasets. Samples overlapping with HMDB51 are excluded from each OOD dataset to maintain distinct ID/OOD classes. For instance, 31 classes overlapping with HMDB51 are removed from UCF101, leaving 70 OOD classes, and 8 overlapping classes are removed from EPIC-Kitchens and HAC.

\noindent
\textbf{Kinetics-600 as ID}: We designate UCF101, EPIC-Kitchens, HAC, and HMDB51 as OOD datasets, excluding any ID class overlap with Kinetics-600. For example, 11 overlapping classes are removed from UCF101, leaving 90 OOD classes, while the original classes in EPIC-Kitchens, HAC, and HMDB51 are preserved as they have no overlap with Kinetics-600.

\subsection{Baseline Design}
\label{appx:baseline}
As briefly discussed in \S \ref{subsec:task_base_imple}, we compare \method against traditional on-device training classifiers combined with various post-hoc OOD detection methods. Additionally, we design two alternative baselines that do not require on-device training to further analyze the effectiveness of our approach.  

For the OOD detection methods, we extend several established techniques to the multimodal setting, including MSP \cite{oodbaseline17iclr}, Energy \cite{energyood20nips}, MaxLogit \cite{hendrycks2019anomalyseg}, Mahalanobis \cite{mahananobis18nips}, ReAct \cite{sun2021tone}, ASH \cite{djurisic2022extremely}, GEN \cite{liu2023gen}, KNN \cite{sun2022knnood}, and VIM \cite{wang2022ofa}. These methods span multiple levels of OOD scoring, ranging from probability-based approaches (MSP, GEN), logit-based techniques (Energy, MaxLogit), and feature-space methods (Mahalanobis, KNN) to activation manipulation strategies (ReAct, ASH) and hybrid logit-feature approaches (VIM). 

For the newly designed baseline methods, Ini-Classifier denotes a randomly initialized classifier that is used directly for inference without any training or fine-tuning. In contrast, Ini-Hypernetwork refers to a randomly initialized Hyper-Network that generates classifier parameters based on extracted visual features.

\subsection{Implementation Details}
\label{appx:im_details}
As briefly discussed in \S \ref{subsec:task_base_imple}, Near-OOD and Far-OOD tasks share the batch size of 16, the Adam optimizer, and the learning rate of 0.0001. In addition, the machine we used in the experiments is as follows:

GPU server with AMD EPYC Milan 7763, 64×16 = 1TB DDR4 memory, 15 TB SSD, 6× NVIDIA RTX A6000 Ada.

The encryption time was measured on a MacBook Pro with M1 Pro.

For the Near-OOD tasks and for Far-OOD tasks with HMDB as the ID dataset (excluding Kinetics), the hypernetwork is configured as a single layer—there is no hidden layer. The weight matrix $W$ of size $(\text{num feature}\times \text{num class})$ and the bias $b$ of size $(\text{num class})$ are directly derived from the input features, with a batch normalization applied before the parameter output.

For the Near-OOD and Far-OOD tasks with Kinetics as the ID dataset, the hypernetwork is configured with two layers, where the hidden layer dimension is set to 3584 for Near-OOD and 2048 for Far-OOD. In this case, the parameters undergo batch normalization at the latent variable stage before output;  the weight matrix $W$ of size $(\text{num feature}\times \text{num class})$ and the bias $b$ of size $(\text{num class})$ are the results after the second batch normalization.
\begin{table}[t]
    \centering
    \renewcommand{\arraystretch}{1.2}
    \setlength{\tabcolsep}{3pt}
    \scriptsize
    \begin{tabular}{l cc cc cc}
        \toprule
        \multirow{2}{*}{\textbf{Methods}} & 
        \multicolumn{2}{c}{\textbf{No Mask}} &
        \multicolumn{2}{c}{\textbf{Mask 50\% Channels}} &
        \multicolumn{2}{c}{\textbf{Mask 75\% Channels}}\\
        \cmidrule(lr){2-3} \cmidrule(lr){4-5} \cmidrule(lr){6-7}
         & FPR95$\downarrow$ & AUROC$\uparrow$ 
         & FPR95$\downarrow$ & AUROC$\uparrow$ 
         & FPR95$\downarrow$ & AUROC$\uparrow$\\
        \midrule
        \multicolumn{7}{c}{\textbf{HMDB as ID}} \\ 
        \midrule
        Kinetics      & 21.89  & 94.29  & 15.39  & 96.06  & 43.79   & 89.97\\
        UCF      & 52.79  & 81.93  & 46.18  & 86.19  & 68.76  & 74.61\\
        HAC  & 29.42  & 94.02  & 22.35  & 94.41  & 29.76  & 93.91\\
        \midrule
        \multicolumn{7}{c}{\textbf{Kinetics as ID}} \\ 
        \midrule
        HMDB      & 69.67  & 78.09  & 69.52  & 79.25  & 70.19   & 75.66\\
        UCF      & 69.63  & 72.98  & 69.34  & 75.37  & 70.89  & 72.84\\
        HAC  & 68.01  & 76.81  & 58.32  & 84.98  & 69.36   & 76.29\\
        \bottomrule
    \end{tabular}
    \caption{Far-OOD Detection results using various mask proportions ($\uparrow$ higher is better; $\downarrow$ lower is better).}
    \label{tab:mask_far}
\end{table}

\subsection{Additional Results}
\label{appx:addi_res}
As discussed in \S \ref{subsec:IDA}, we present additional experimental results to further analyze the impact of different masking percentages. 

Tab.~\ref{tab:mask_far} summarizes the model’s performance under various masking conditions.
Interestingly, we observe that applying a 50\% masking ratio results in only a slight performance degradation. This finding suggests that our dynamic channel sampling approach effectively removes noisy or redundant channels while maintaining the model’s overall capability. We attribute this robustness to the efficiency of our sampling strategy, which dynamically selects the most informative channels, thereby mitigating the negative impact of extensive masking. Furthermore, this result provides additional evidence for the resilience of the hypernetwork, as it demonstrates that the model does not rely on highly specific input features to perform well.
\begin{table}[t]
    \renewcommand{\arraystretch}{1.2}
    \setlength{\tabcolsep}{3pt}
    \scriptsize
    \centering
    \label{tab:time-delay-updated}
    \begin{tabular}{L{1cm}|P{1.2cm}|P{1.1cm}|P{1.2cm}|P{1.2cm}|P{1.2cm}}
    \toprule
    \textbf{Datasets} & \textbf{Size} & \textbf{4G: 5MB/s} & \textbf{4G: 15MB/s} & \textbf{5G: 50MB/s} & \textbf{5G: 100MB/s} \\
    \midrule
    \multicolumn{6}{c}{\textbf{Near-OOD}} \\
    \midrule
    HMDB &
    \makecell[l]{%
      \mystrut $\uparrow:$ 0.18\,MB\\
      \mystrut $\downarrow:$ 4.75\,MB%
      }
    &
    \makecell[l]{%
      \mystrut $\uparrow:$ 144\,ms\\
      \mystrut $\downarrow:$ 950\,ms%
      }
    &
    \makecell[l]{%
      \mystrut $\uparrow:$ 48\,ms\\
      \mystrut $\downarrow:$ 317\,ms%
        }
    &
    \makecell[l]{%
      \mystrut $\uparrow:$ 14\,ms\\
      \mystrut $\downarrow:$ 95\,ms%
        }
    &
    \makecell[l]{%
      \mystrut $\uparrow:$ 7\,ms\\
      \mystrut $\downarrow:$ 48\,ms%
        }
    \\
    \midrule
    UCF &
    \makecell[l]{%
      \mystrut $\uparrow:$ 0.18\,MB\\
      \mystrut $\downarrow:$ 9.40\,MB%
        }
    &
    \makecell[l]{%
      \mystrut $\uparrow:$ 144\,ms\\
      \mystrut $\downarrow:$ 1880\,ms%
        }
    &
    \makecell[l]{%
      \mystrut $\uparrow:$ 48\,ms\\
      \mystrut $\downarrow:$ 627\,ms%
        }
    &
    \makecell[l]{%
      \mystrut $\uparrow:$ 14\,ms\\
      \mystrut $\downarrow:$ 188\,ms%
        }
    &
    \makecell[l]{%
      \mystrut $\uparrow:$ 7\,ms\\
      \mystrut $\downarrow:$ 94\,ms%
        }
    \\
    \midrule
    Kinetics &
    \makecell[l]{%
      \mystrut $\uparrow:$ 0.18\,MB\\
      \mystrut $\downarrow:$ 24.17\,MB%
    }
    &
    \makecell[l]{%
      \mystrut $\uparrow:$ 144\,ms\\
      \mystrut $\downarrow:$ 4834\,ms%
    }
    &
    \makecell[l]{%
      \mystrut $\uparrow:$ 48\,ms\\
      \mystrut $\downarrow:$ 1611\,ms%
    }
    &
    \makecell[l]{%
      \mystrut $\uparrow:$ 14\,ms\\
      \mystrut $\downarrow:$ 483\,ms%
    }
    &
    \makecell[l]{%
      \mystrut $\uparrow:$ 7\,ms\\
      \mystrut $\downarrow:$ 242\,ms%
    }
    \\
    \midrule
    EPIC &
    \makecell[l]{%
      \mystrut $\uparrow:$ 0.18\,MB\\
      \mystrut $\downarrow:$ 0.79\,MB%
    }
    &
    \makecell[l]{%
      \mystrut $\uparrow:$ 144\,ms\\
      \mystrut $\downarrow:$ 158\,ms%
    }
    &
    \makecell[l]{%
      \mystrut $\uparrow:$ 48\,ms\\
      \mystrut $\downarrow:$ 53\,ms%
    }
    &
    \makecell[l]{%
      \mystrut $\uparrow:$ 14\,ms\\
      \mystrut $\downarrow:$ 16\,ms%
    }
    &
    \makecell[l]{%
      \mystrut $\uparrow:$ 7\,ms\\
      \mystrut $\downarrow:$ 8\,ms%
    }
    \\
    \midrule
    \multicolumn{6}{c}{\textbf{Far-OOD}} \\
    \midrule
    HMDB &
    \makecell[l]{%
      \mystrut $\uparrow:$ 0.18\,MB\\
      \mystrut $\downarrow:$ 8.08\,MB%
    }
    &
    \makecell[l]{%
      \mystrut $\uparrow:$ 144\,ms\\
      \mystrut $\downarrow:$ 1616\,ms%
    }
    &
    \makecell[l]{%
      \mystrut $\uparrow:$ 48\,ms\\
      \mystrut $\downarrow:$ 539\,ms%
    }
    &
    \makecell[l]{%
      \mystrut $\uparrow:$ 14\,ms\\
      \mystrut $\downarrow:$ 162\,ms%
    }
    &
    \makecell[l]{%
      \mystrut $\uparrow:$ 7\,ms\\
      \mystrut $\downarrow:$ 81\,ms%
    }
    \\
    \midrule
    Kinetics &
    \makecell[l]{%
      \mystrut $\uparrow:$ 0.18\,MB\\
      \mystrut $\downarrow:$ 42.81\,MB%
    }
    &
    \makecell[l]{%
      \mystrut $\uparrow:$ 144\,ms\\
      \mystrut $\downarrow:$ 8562\,ms%
    }
    &
    \makecell[l]{%
      \mystrut $\uparrow:$ 48\,ms\\
      \mystrut $\downarrow:$ 2854\,ms%
    }
    &
    \makecell[l]{%
      \mystrut $\uparrow:$ 14\,ms\\
      \mystrut $\downarrow:$ 856\,ms%
    }
    &
    \makecell[l]{%
      \mystrut $\uparrow:$ 7\,ms\\
      \mystrut $\downarrow:$ 428\,ms%
    }
    \\
    \bottomrule
    \end{tabular}
    \caption{Time delay with updated encryption/decryption method and separated upload/download speeds. (\(\uparrow\)) denotes upload from device to cloud, (\(\downarrow\)) denotes download from cloud to device. The upload speed is assumed to be 1/4 of the download speed.}
    \label{tab:time-delay-updated}
\end{table}

Tab.~\ref{tab:time-delay-updated} presents the communication latency for different datasets under various network conditions, considering an updated encryption/decryption method. The upload speed is assumed to be one-fourth of the download speed, reflecting typical mobile network conditions. The results show that 4G networks with a 5MB/s download speed incur the highest delays, especially for large datasets such as Kinetics, where the download time reaches 8.56 seconds. In contrast, 5G networks significantly reduce latency, with the fastest configuration (100MB/s) achieving sub-500ms downloads for most cases. The encryption/decryption overhead is minimal, as observed in the slight increase in total transmission time. These findings highlight the benefits of high-speed networks for cloud-assisted OOD detection, particularly in handling large datasets efficiently.

\end{document}